\documentclass[aps,pra,twocolumn,showpacs,amsmath,amssymb]{revtex4}
\usepackage{bm,mathrsfs}
\usepackage{hyperref}
\renewcommand{\vec}[1]{\bm{#1}}
\renewcommand{\asymp}{\sim}
\renewcommand{\imath}{i}

\begin{document}

\title{Levinson theorem for Aharonov--Bohm scattering in two dimensions}

\author{Denis D. Sheka}\thanks{Alexander von Humboldt Fellow}
\affiliation{National Taras Shevchenko University of Kiev, 03127 Kiev,
Ukraine}
\email{Denis_Sheka@univ.kiev.ua} %

\author{Franz G. Mertens}
\affiliation{Physikalisches Institut, Universit\"at Bayreuth, D--95440
Bayreuth, Germany}

\date{September 25, 2006}

\begin{abstract}
We apply the recently generalized Levinson theorem for potentials with inverse
square singularities [Sheka \emph{et al}, Phys.Rev.A \textbf{68}, 012707
(2003)] to Aharonov--Bohm systems in two--dimensions. By this theorem, the
number of bound states in a given $m$--th partial wave is related to the phase
shift and the magnetic flux. The results are applied to 2D soliton--magnon
scattering.
\end{abstract}

%
% 03.65.Nk    Scattering theory
% 73.50.Bk    General theory, scattering mechanisms
% 03.65.Vf    Phases: geometric; dynamic or topological
% 75.10.Hk    Classical spin models

\maketitle

\section{Introduction}

In \citeyear{Levinson49} \citet{Levinson49} established one of the most
beautiful results of the scattering theory: the Levinson theorem sets up a
relation between the number of bound states $N_l^{\text{b}}$ in a given $l$-th
partial wave and the phase shift $\delta_l(k)$, namely
$\delta_l(0)-\delta_l(\infty) = \pi N_l^{\text{b}}$.

Ten years later, in \citeyear{Aharonov59}, \citet{Aharonov59} discovered the
global properties of the magnetic flux. Nowadays the Aharonov--Bohm (AB)
effect is often involved to understand different quantum--mechanical phenomena
\cite{Peshkin89}.

The aim of this paper is to generalize the Levinson theorem to systems which
exhibit AB effects. We denote such systems as AB--systems. Recently, the
analogue of the Levinson theorem was established by \citet{Lin03} for the
simplest AB--system with constant magnetic flux $\Phi$. Here we establish a
more general relation, valid for a magnetic field with a vector potential of
the form
\begin{equation} \label{eq:A-defn}
\begin{split}
\vec{A}(\vec{r}) &= \frac{\Phi(\rho)}{2\pi}\vec{\nabla}\chi
\equiv \frac{\Phi(\rho)}{2\pi\rho}\,\vec{e}_\chi,\\
\Phi(0)& = 2\pi\alpha, \qquad \Phi(\infty) = 2\pi\beta.
\end{split}
\end{equation}
Here $\rho$ and $\chi$ are the polar coordinates in two spatial dimensions.

The paper is organized as follows. In Sec.~\ref{sec:notations} we formulate
the scattering problem for the AB--systems \eqref{eq:A-defn}. We proof the
Levinson theorem for the simplest so--called \emph{centrifugal} AB--model in
Sec.~\ref{sec:centrifugal}. The general form of the theorem is established in
Sec.~\ref{sec:proof}. We compare our results with those for the conventional
AB system in Sec.~\ref{sec:discussion} and discuss the physical meaning of the
extra term in the generalized theorem. In Sec.~\ref{sec:application} we apply
our results to 2D magnetic systems. Namely, we consider the soliton--magnon
scattering, which can be described in the framework of AB--scattering of the
general form \eqref{eq:A-defn}. Concluding remarks are presented in
Sec.~\ref{sec:conclusion}.

\section{Scattering problem for the AB--system: notations, partial wave expansion}
\label{sec:notations} %

Let us consider the Schr\"{o}dinger--like equation for a spinless particle in
a magnetic field in two dimensions:
\begin{equation} \label{eq:Schroedinger}
\left(-\imath\vec{\nabla} - \vec{A}\right)^2\Psi + V(\vec{r})\Psi=\imath
\partial_t\Psi.
\end{equation}

We will consider a central (axially symmetric) potential, $V(\vec{r})=V(\rho)$
and a magnetic vector potential in the form \eqref{eq:A-defn}. Such a form of
the magnetic field is typical for the Aharonov--Bohm effect, it corresponds to
the magnetic induction
\begin{equation*}
\vec{B} = \vec{\nabla} \times \vec{A} = \vec{e}_z \left[\frac{\Phi'(\rho
)}{2\pi\rho} + \Phi(\rho )\, \delta(\vec{r}) \right].
\end{equation*}
Thus, the magnetic field has a singular point at the origin (vortex line). The
total magnetic flux is $\int B_z \mathrm{d}^2x = \Phi(\infty)$.

We will denote the systems with above mentioned potentials as AB--systems. For
such systems it is possible to apply the standard partial wave expansion,
using the \emph{ansatz}
\begin{equation} \label{eq:ansatz}
\Psi(\vec{r},t)=\sum_{m=-\infty}^\infty\psi_m^{\mathcal{E}}(\rho)\cdot
\exp(\imath m\chi-\imath \mathcal{E}t)\;,
\end{equation}
where $\{m,\mathcal{E}\}$ is the complete set of eigennumbers, $\mathcal{E}$
and $m$ are the energy and the azimuthal quantum number, respectively. Each
partial wave $\psi_m^{\mathcal{E}}$ is an eigenfunction of the spectral
problem
\begin{subequations} \label{eq:EVP&Um}
\begin{equation} \label{eq:EVP}
H\psi_m^{\mathcal{E}}(\rho) = \mathcal{E}\psi_m^{\mathcal{E}}(\rho)
\end{equation}
for the 2D radial Schr\"{o}dinger operator $H=-\nabla_\rho^2+U_m(\rho)$ with
the partial potential
\begin{equation} \label{eq:Um}
U_m(\rho) = V(\rho)+\frac{\left[ m -
\dfrac{\Phi(\rho)}{2\pi}\right]^2}{\rho^2}.
\end{equation}
\end{subequations}

Let us formulate the scattering problem. A continuum spectrum exists for
$\mathcal{E}>0$. Note that the eigenfunctions for the free particle,
$V(\rho)=\Phi(\rho)= 0$, have the form
\begin{equation} \label{eq:psi-free}
\psi_m^{\text{free}}(\rho)\propto J_m(k\rho), \qquad k =
\sqrt{\mathcal{E}}>0\;,
\end{equation}
where $k$ is a ``radial wave number'', and $J_m$ is a Bessel function. The
free eigenfunctions like $\psi_m^{\text{free}}$ play the role of partial
cylinder waves of the plane wave
\begin{equation} \label{eq:plane-wave}
\exp(\imath \vec{k}\cdot\vec{r}-\imath \mathcal{E} t) =
\sum_{m=-\infty}^\infty \imath^m J_m(k\rho)e^{\imath m\chi-\imath
\mathcal{E}t}.
\end{equation}

The behavior of the eigenfunctions in the potentials $V(\rho)$ and $\Phi(\rho
)$ can be analyzed at large distances from the origin, $\rho\gg R$, where $R$
is a typical range of the potentials. In view of the asymptotic behavior
$U_m(\rho)\asymp m^2/\rho^2$, which is valid for fast decreasing potentials
$V(\rho)$ and $\Phi (\rho )$, in the leading approximation in $1/\rho$ we have
the usual result
\begin{equation} \label{eq:psi-scat}
\psi_m^{\mathcal{E}} \propto J_{|m|}(k\rho) + \sigma_m(k)Y_{|m|}(k\rho)\;,
\end{equation}
where $Y_m$ is a Neumann function. The quantity $\sigma_m(k)$ stems from the
scattering; it can be interpreted as the scattering amplitude. In the limiting
case $k\rho\gg |m|$ it is convenient to consider the asymptotic form of
Eq.~\eqref{eq:psi-scat},
\begin{equation*}
\psi_m^{\mathcal{E}} \propto \frac{1}{\sqrt{\rho}}\cos\left(k\rho -
\frac{|m|\pi}{2} - \frac{\pi}{4}+\delta_m(k)\right),
\end{equation*}
where the scattering phase, or the phase shift
$\delta_m(k)=-\arctan\sigma_m(k)$. The phase shift contains all informations
about the scattering process. In particular, we give the general solution of
the scattering problem for the plane wave \eqref{eq:plane-wave}. With
Eqs.~\eqref{eq:ansatz} and \eqref{eq:psi-scat}, the asymptotic solution of the
Schr\"{o}dinger--like equation \eqref{eq:Schroedinger} for $\rho\gg R$ can be
written as
\begin{equation} \label{eq:Psi-general}
\begin{split}
\Psi(\vec{r},t)&=\sum_{m=-\infty}^\infty C_m \left[J_{|m|}(k\rho) +
\sigma_m(k)Y_{|m|}(k\rho)\right]\\
&\times \exp(\imath m\chi - \imath \mathcal{E}t)\;,
\end{split}
\end{equation}
where $C_m$ are constants. To solve the scattering problem for the plane wave
let us choose the constants $C_m$ by comparing Eq.~\eqref{eq:Psi-general} with
the expansion \eqref{eq:psi-free} for the free motion.  Using the asymptotic
forms for the cylinder functions in the region $\rho\gg 1/k$, we obtain
\begin{equation} \label{eq:Psi-scat-via-F}
\begin{split}
\Psi(\vec{r},t) &=e^{\imath \vec{k}\cdot\vec{r}-\imath \mathcal{E}t}
+\mathcal{F}(\chi)\frac{e^{\imath k\rho-\imath \mathcal{E}t}}{\sqrt{\rho}},\\
\mathcal{F}(\chi) &= \frac{\exp(-\imath \pi/4)}{\sqrt{2\pi
k}}\cdot\sum_{m=-\infty}^{\infty} \left(e^{2\imath \delta_m}-1\right)\cdot
e^{\imath m\chi}.
\end{split}
\end{equation}
The total scattering cross section is given by the expression
\begin{equation} \label{eq:total}
\mathfrak{S}^{\text{tot}} = \int_0^{2\pi}\!|\mathcal{F}|^2d\chi =
\sum_{m=-\infty}^\infty\! \mathfrak{S}_m\;,
\end{equation}
where $\mathfrak{S}_m =(4/k)\sin^2\delta_m$ are the partial scattering cross
sections.

\section{Levinson theorem for the AB--model} %
\label{sec:AB} %

For regular 2D potentials $V(\rho)$ without magnetic field [$\Phi(\rho)=0]$,
the 2D analogue of the Levinson theorem has the form
\cite{Bolle86,Lin97,Dong98}
\begin{equation} \label{eq:Levinson}
\delta_m(0)-\delta_m(\infty) = \pi\cdot \left(
N_m^{\text{b}}+N_m^{\text{hb}}\cdot\delta_{|m|,1}\right).
\end{equation}
Here $N_m^{\text{b}}$ is the number of bound states in a given $m$--th partial
wave, $N_m^{\text{hb}}$ is the number of half bound states (recall that a
zero--energy state is called a half bound state if its wave function is
finite, but does not decay fast enough at infinity to be square integrable).

Here all partial potentials $U_m(\rho)$ satisfy the asymptotic conditions
\begin{subequations} \label{eq:as08}
\begin{align} \label{eq:as0}
\lim_{\rho=0}\rho^2U_m(\rho)&=m^2,\\ %
\label{eq:as8} %
\lim_{\rho=\infty}\rho^2U_m(\rho)&=m^2,
\end{align}
\end{subequations}
which provide a regular behavior at the origin, and fast decaying at infinity.

The presence of the nonlocal magnetic field can break the asymptotic
conditions \eqref{eq:as08}. Namely, if the field does not vanish at the
origin, $\Phi(0) = 2\pi \alpha  \neq0$, the asymptotic condition
\eqref{eq:as0} is broken. In the same way, a not vanishing field at infinity,
$\Phi(\infty) = 2\pi \beta \neq0$, breaks the asymptotic condition
\eqref{eq:as8}. There appear inverse square singularities in the effective
partial potential at origin or in the inverse square tail at infinity. The
standard Levinson theorem fails for this case \cite{Sheka03,Lin03} and some
generalization is needed.

Before we discuss the general case, let us consider the simplest AB--model,
which nevertheless contains the main features of the problem.

\subsection{The simplest ``centrifugal'' AB--model} %
\label{sec:centrifugal} %

We start with vector potentials of the form
\begin{equation} \label{eq:centrifugal}
\vec{A}(\vec{r}) =
  \begin{cases}
    \alpha\vec{\nabla}\chi & \text{when $\rho < R$}, \\
    \beta\vec{\nabla}\chi  & \text{otherwise},
  \end{cases}
\end{equation}
where $\alpha$ and $\beta$ are nonzero constants. For this simple model the
potential $V(\rho)\equiv0$, so the effective partial potentials \eqref{eq:Um}
for the correspondent spectral problem \eqref{eq:EVP&Um} can be rewritten as
follows:
\begin{align*} \label{eq:U_m-cf}
U_m(\rho) &=
  \begin{cases}
    \dfrac{\nu^2}{\rho^2} & \text{when $\rho < R$}, \\
    \dfrac{\mu^2}{\rho^2}  & \text{otherwise},
  \end{cases}\\
%\label{eq:nu&mu}%
\nu&\equiv m - \alpha, \qquad \mu\equiv m - \beta.
\end{align*}
The scattering problem for this so--called \emph{centrifugal model} has an
exact solution, see \cite{Sheka03}:
\begin{equation*} \label{eq:delta-centrifugal}
\begin{split}
\delta_m^{\text{cf}}(k) &= \frac{|m|-|\mu|}{2}\cdot\pi - \arctan
{\tilde{\sigma}}_{\mu}^{\text{cf}}(\varkappa\equiv kR)\;,\\
{\tilde{\sigma}}_{\mu}^{\text{cf}}(\varkappa) &= \frac{
J_{|\nu|}^{\prime}(\varkappa)\cdot J_{|\mu|}(\varkappa)-
J_{|\mu|}^{\prime}(\varkappa)\cdot J_{|\nu|}(\varkappa)}
{J_{|\nu|}(\varkappa)\cdot Y_{|\mu|}^{\prime}(\varkappa)-
J_{|\nu|}^{\prime}(\varkappa)\cdot Y_{|\mu|}(\varkappa)}\;.
\end{split}
\end{equation*}
Using the asymptotic form of the cylinder functions, one can easily derive the
Levinson relation for the centrifugal model \cite{Sheka03}:
\begin{equation} \label{eq:Levinson-centrifugal}
\delta_m^{\text{cf}}(0) - \delta_m^{\text{cf}}(\infty) = \frac{\pi}{2}\left(
|\nu|-|\mu|\right).
\end{equation}

As an example we consider the solenoid of zero radius with the constant
magnetic flux $\Phi_0$ and returned flux uniformly distributed on the surface
of a cylinder at radius $R$; the vector potential of such a system is
\cite{Kobe88,Decanini03}
\begin{equation} \label{eq:A-returned}
\vec{A} =
  \begin{cases}
    \dfrac{\Phi_0}{2\pi\rho}\vec{e}_\chi & \text{when $\rho < R$}, \\
    0 & \text{otherwise}.
  \end{cases}
\end{equation}
The magnetic induction $\vec{B} = \frac{\Phi_0}{2\pi\rho}\bigl[\delta(\rho) -
\delta(\rho-R)\bigr]\vec{e}_z$ consists of the usual AB flux line at $\rho=0$
and an infinitely thin magnetic field shell at $\rho=R$. Identifying
parameters $\alpha=\Phi_0/2\pi$ and $\beta=0$, one can rewrite the Levinson
relation \eqref{eq:Levinson-centrifugal} as follows:
\begin{equation} \label{eq:Levinson-centrifugal-example}
\delta_m(0) - \delta_m(\infty) = \frac\pi2\left( |m-\Phi_0/2\pi|-|m|\right).
\end{equation}
Note that the Levinson relation takes nonzero values for any nonvanishing AB
field flux $\Phi_0$, which can take also an integer value. In particular, if
$m>\Phi_0/2\pi>0$, the Levinson relation is equal to $-\Phi_0/4$.

\subsection{Levinson theorem for general AB--systems} %
\label{sec:proof} %

Let us discuss the case of the general AB--system with the vector potential of
the form \eqref{eq:A-defn}. We suppose that the particle potential $V(\rho)$
is less singular than $\rho^{-2}$ at origin and decays faster than $\rho^{-2}$
at infinity. Then the partial potential \eqref{eq:Um} satisfies the asymptotic
conditions
\begin{equation} \label{eq:U-as}
U_m(\rho) \asymp
  \begin{cases}
    \dfrac{\nu^2}{\rho^2}, & \text{when $\rho\to0$}, \\
    \dfrac{\mu^2}{\rho^2}, & \text{when $\rho\to\infty$},
  \end{cases}
\end{equation}
where $\nu= m - \alpha$ and $\mu = m - \beta$. In the presence of magnetic
flux at least one of the parameters $\alpha$ and $\beta$ has a nonzero value.
This breaks the regular asymptotic conditions \eqref{eq:as08}. In the general
case there appears an effective potential, which has an inverse square
singularity at the origin ($\nu\neq m$) and an inverse square tail at infinity
($\mu\neq m$). The Levinson theorem for such singular potentials was
generalized in our recent paper \cite{Sheka03}. Namely, when an effective
partial potential has the asymptotic behavior \eqref{eq:U-as}, the generalized
Levinson theorem \cite{Sheka03} reads:
\begin{equation} \label{eq:Levinson-gen}
\delta_m(0) - \delta_m(\infty) = \pi\cdot \left( N_m^{\text{b}}+
\frac{|\nu|-|\mu|}{2}\right).
\end{equation}
Identifying the parameters $\nu$ and $\mu$ one can rewrite the Levinson
relation in the following form:
\begin{equation} \label{eq:AB-Levinson}
\delta_m(0) - \delta_m(\infty) = \pi\cdot \left( N_m^{\text{b}}+ \frac{|m -
\alpha| - |m - \beta|}{2}\right).
\end{equation}

\section{Discussion}
\label{sec:discussion} %

The Levinson theorem \eqref{eq:AB-Levinson} establishes a relation between the
number of bound states in a given $m$--th partial wave, the total phase shift,
and the magnetic flux.

Let us discuss the physical meaning of the extra term
\begin{equation} \label{eq:extra}
\frac{\pi}{2}\left(|m - \alpha| - |m - \beta|\right)
\end{equation}
in the Levinson relation. This term results from the long--range behavior of
the AB potential. Singular behavior of the AB potential at origin creates a
``vorticity'' $\alpha$, which induces wave functions with $m$ greater
(smaller) than $\alpha$ to go around the origin in the counterclockwise
(clockwise) direction. Thus the short--wavelength scattering data are shifted
by $(\pi/2)(|m|-|m-\alpha|)$. The same situation takes place for AB potentials
with long--range tail, which creates a ``vorticity'' $\beta$, and the
long--wavelength scattering data are changed by $(\pi/2)(|m-\beta|-|m|)$. As a
result, the correction to the Levinson relation takes the form
\eqref{eq:extra}.

Let us compare our results with those for the conventional AB system
\cite{Aharonov59,Aharonov84}:
\begin{equation} \label{eq:AB-conventional}
\vec{A} =
  \begin{cases}
    \frac12 B\rho\, \vec{e}_\chi & \text{when $\rho<R$}, \\
    \dfrac{BR^2}{2\rho}\vec{e}_\chi & \text{otherwise}.
  \end{cases}
\end{equation}
Such a field produces a constant magnetic induction $\vec{B}=B\vec{e}_z$
inside the cylinder of radius $R$, and provides an empty induction outside.
The Levinson relation for this case reads
\begin{equation*} \label{eq:Levinson4AB-standard}
\delta_m(0) - \delta_m(\infty) = \frac{\pi}{2}\left(|m|- |m-\beta|\right), \;
\beta = \tfrac12 BR^2,
\end{equation*}
in agreement with exact results \cite{Henneberger80,Ruijsenaars83}.

Let us remind that the AB total scattering cross section vanishes when $\beta
\in\mathbb{Z}$. Nevertheless any AB field changes the standard Levinson
relation, even when $\beta \in\mathbb{Z}$. Due to the non--locality of AB
potentials, the total phase shifts do not go to zero with increasing $|m|$. To
treat such a singularity the regularization is usually involved
\cite{Henneberger80} to determine the total scattering amplitude
\eqref{eq:Psi-scat-via-F} or the total scattering cross section
\eqref{eq:total}. The same picture takes place not only for the conventional
AB system \eqref{eq:AB-conventional}, but also for the general case
\eqref{eq:A-defn}. An exception is a simple AB--system with
\begin{equation} \label{eq:A-simple}
\vec{A}(\vec{r})=\alpha\vec{\nabla}\chi.
\end{equation}
In this case one has a standard Levinson relation in the form $ \delta_m(0) -
\delta_m(\infty) = \pi N_m^{\text{b}}.$ One should note that nevertheless each
scattering state $\delta_m(k)$ corresponds to a given general angular momentum
$\nu=|m-\alpha|$. The Levinson theorem for this particular case was first
obtained by \citet{Lin03}.

The Levinson relations should be modified for the critical case when half
bound states occur. The Levinson theorem for the system with possible
half--bound states has been considered first by \citet{Bolle86}, and
reestablished later by another method by \citet{Dong98}. Without magnetic
field the Levinson relation has the form of Eq.~\eqref{eq:Levinson}, so the
half--bound states affect in the same way the two modes with $m=\pm1$. The
presence of the magnetic field breaks the symmetry
$\delta_m(k)=\delta_{-m}(k)$, and in the general case the contribution of the
half--bound states in the form \eqref{eq:Levinson} can not be adequate.
However, for the particular case \eqref{eq:A-simple}, the problem can be
solved \cite{Lin03}.

If the particle potential $V(\rho)$ has an inverse square singularity, or an
inverse square tail, then the Levinson theorem in the form
\eqref{eq:AB-Levinson} fails. Instead, one has to calculate the effective
intensities $\nu$ and $\mu$ of the singularities in the partial potential as
follows
\begin{equation} \label{eq:as08-AB}
\nu^2=\lim_{\rho=0}\rho^2U_m(\rho),\qquad
\mu^2=\lim_{\rho=\infty}\rho^2U_m(\rho)
\end{equation}
and then one obtains  the Levinson theorem in the form
\eqref{eq:Levinson-gen}.

\section{Applications to magnetism: scattering
on a magnetic soliton in 2D isotropic magnets} %
\label{sec:application} %

All mentioned above results can be applied to a wide class of AB--systems.
Here we do not consider a quantum--mechanical example of the general
AB--scattering system. Namely, we apply our results to the description of the
soliton--magnon interaction in a 2D magnet. Note that it is possible to apply
the quantum AB theorem to a classical system, because magnons in a magnet can
be formally described by a Scr\"odinger--like equation with an effective
magnetic field in the form which is typical for AB-systems.

We consider the model of the 2D isotropic Heisenberg ferromagnet, where the
elementary linear excitations of the spin system (magnons) can coexist
together with nonlinear ones (solitons). In terms of the angular variables for
the normalized magnetization $\vec{m} = \left( \sin\theta\cos\phi;
\sin\theta\sin\phi; \cos\theta\right)$, the structure of the simplest
nonlinear excitation, the so--called Belavin--Polyakov soliton is described by
the formulae \cite{Belavin75}
\begin{equation*} \label{eq:BP-soliton}
\tan \frac{\theta_0(\rho)}{2} = \left(\frac{R}{\rho}\right)^{|q|}, \qquad
\phi_0 = \varphi_0 + q\chi.
\end{equation*}
Here $q\in\mathbb{Z}$ is the topological charge of the soliton, $R$ and
$\varphi_0$ are arbitrary parameters.

To analyze the soliton--magnon interaction, one considers small oscillations
of the magnetization $(\theta,\phi)$ on the background of the soliton
$(\theta_0, \phi_0)$. These oscillations can be described in terms of the
complex valued ``wave function'' $\psi = \theta-\theta_0 + \imath \sin\theta_0
(\phi-\phi_0)$. The linearized equations have the form of the
Schr\"{o}dinger--like equation \eqref{eq:Schroedinger} with an effective
potential \cite{Ivanov99,Ivanov05b}
\begin{equation*} \label{eq:V-FM}
V(\rho) = -\frac{q^2}{\rho^2}\sin^2\theta_0
\end{equation*}
and an effective magnetic field in the form
\begin{align*}
\vec{A}(\vec{r}) &= \frac{\Phi(\rho)}{2\pi}\vec{\nabla}\chi, \quad
\Phi(\rho) = -2\pi q\cos\theta_0(\rho) \\
\Phi(0)& = 2\pi q,  \quad \Phi(\infty) = -2\pi q.
\end{align*}
The partial potential \eqref{eq:Um} has the form \cite{Ivanov99}
\begin{equation*} \label{eq:Um-FM}
U_m(\rho) = \frac{m^2 + 2mq\cos\theta_0(\rho) +
q^2\cos2\theta_0(\rho)}{\rho^2}.
\end{equation*}
Using Eq.~\eqref{eq:as08-AB}, one can calculate the intensities of the inverse
square singularities:
\begin{equation*} \label{eq:nu&nu-FM}
\nu = |m-q|, \qquad \mu= |m+q|.
\end{equation*}
The Levinson theorem reads
\begin{equation} \label{eq:Levinson-FM}
\delta_m(0) - \delta_m(\infty) = \pi\left( N_m^{\text{b}} + N_m^{\text{hb}} +
\frac{|m-q| - |m+q|}{2}\right).
\end{equation}
As found by \citet{Ivanov95g}, the soliton with a topological charge $q$ has
$2|q|$ internal zero frequency modes, when $m\in[-q+1;q]$. Namely, modes with
$m\in[-q+2;q]$ form bound state, while the mode with $m=-q+1$ is the
half--bound states. Finally the Levinson theorem for the soliton--magnon
scattering takes the form (we chose $q>0$)
\begin{equation*} \label{eq:Levinson-FM1}
\frac{\delta_m(0) - \delta_m(\infty)}{\pi} =
  \begin{cases}
    q   & \text{when $m\leq -q$}, \\
    1-m & \text{when $-q<m\leq q$},\\
    -q  & \text{when $m> q$}.
  \end{cases}
\end{equation*}
This result agrees with our previous analytical and numerical calculation for
the soliton with $q=1$, see Ref.~\cite{Ivanov99}.

Note that the phase shift varies in a wide range, so it \emph{can not be
described}, not even approximately, in the framework of the Born approximation. It
was the source of numerous inconsistencies between previous attempts to
calculate the soliton--magnon interaction in magnets
\cite{Rodriguez89,Pereira95,Ivanov99a}. The reason is that due to the
non--locality of the AB magnetic field, the perturbative Born approximation is
not adequate for the AB scattering \cite{Feinberg63,Henneberger80}.

\section{Conclusion} %
\label{sec:conclusion} %

In conclusion, we have applied our recent results \cite{Sheka03} for the
scattering in a singular potential to AB--systems, and established a
generalization of the Levinson theorem. The theorem constructs the relation
between the number of bound states $N_m^{\text{b}}$ in a given $m$--th partial
wave, the total phase shift $\delta_m(0) - \delta_m(\infty)$ of the scattering
state, and the magnetic flux $\Phi$. When the magnetic flux parameter takes
different values at origin and at infinity, $\alpha=\Phi(0)/2\pi$ and
$\beta=\Phi(\infty)/2\pi$, the Levinson relation takes the form of
Eq.~\eqref{eq:AB-Levinson}. The total phase shift can be treated as a counter
for the bound states.

The generalized Levinson theorem \eqref{eq:AB-Levinson} can be applied to
different AB--systems, including quantum Hall systems, superconductors, and so
forth \cite{Peshkin89}. The method can be used not only for quantum mechanical
AB--systems. In particular, we have verified the theorem for the case of the
soliton--magnon interaction in the 2D isotropic Heisenberg model.

\begin{acknowledgments}
Authors thank Prof. B.~A.~Ivanov for helpful discussions. D.~D.~Sheka thanks
the University of Bayreuth, where part of this work was performed, for kind
hospitality and acknowledges support from the Alexander von Humboldt
Foundation.
\end{acknowledgments}

%\bibliography{soliton} %

\begin{thebibliography}{21}
\expandafter\ifx\csname natexlab\endcsname\relax\def\natexlab#1{#1}\fi
\expandafter\ifx\csname bibnamefont\endcsname\relax
  \def\bibnamefont#1{#1}\fi
\expandafter\ifx\csname bibfnamefont\endcsname\relax
  \def\bibfnamefont#1{#1}\fi
\expandafter\ifx\csname citenamefont\endcsname\relax
  \def\citenamefont#1{#1}\fi
\expandafter\ifx\csname url\endcsname\relax
  \def\url#1{\texttt{#1}}\fi
\expandafter\ifx\csname urlprefix\endcsname\relax\def\urlprefix{URL }\fi
\providecommand{\bibinfo}[2]{#2} \providecommand{\eprint}[2][]{\url{#2}}

\bibitem[{\citenamefont{Levinson}(1949)}]{Levinson49}
\bibinfo{author}{\bibfnamefont{N.}~\bibnamefont{Levinson}},
  \bibinfo{journal}{Selsk. Mat. Fys. Medd.} \textbf{\bibinfo{volume}{25}},
  \bibinfo{pages}{9} (\bibinfo{year}{1949}).

\bibitem[{\citenamefont{Aharonov and Bohm}(1959)}]{Aharonov59}
\bibinfo{author}{\bibfnamefont{Y.}~\bibnamefont{Aharonov}} \bibnamefont{and}
  \bibinfo{author}{\bibfnamefont{D.}~\bibnamefont{Bohm}},
  \bibinfo{journal}{Physical Review} \textbf{\bibinfo{volume}{115}},
  \bibinfo{pages}{485} (\bibinfo{year}{1959}),
  \urlprefix\url{http://prola.aps.org/abstract/PR/v115/i3/p485_1}.

\bibitem[{\citenamefont{Peshkin and Tonomura}(1989)}]{Peshkin89}
\bibinfo{author}{\bibfnamefont{M.}~\bibnamefont{Peshkin}} \bibnamefont{and}
  \bibinfo{author}{\bibfnamefont{A.}~\bibnamefont{Tonomura}}, in
  \emph{\bibinfo{booktitle}{Lecture notes in Physics}}
  (\bibinfo{publisher}{Springer--Verlag}, \bibinfo{address}{Berlin},
  \bibinfo{year}{1989}).

\bibitem[{\citenamefont{Lin}(2003)}]{Lin03}
\bibinfo{author}{\bibfnamefont{D.-H.} \bibnamefont{Lin}},
  \bibinfo{journal}{Phys. Rev. A} \textbf{\bibinfo{volume}{68}},
  \bibinfo{pages}{052705} (\bibinfo{year}{2003}).

\bibitem[{\citenamefont{Boll{\'e} et~al.}(1986)\citenamefont{Boll{\'e},
  Gesztesy, Danneels, and Wilk}}]{Bolle86}
\bibinfo{author}{\bibfnamefont{D.}~\bibnamefont{Boll{\'e}}},
  \bibinfo{author}{\bibfnamefont{F.}~\bibnamefont{Gesztesy}},
  \bibinfo{author}{\bibfnamefont{C.}~\bibnamefont{Danneels}}, \bibnamefont{and}
  \bibinfo{author}{\bibfnamefont{S.~F.~J.} \bibnamefont{Wilk}},
  \bibinfo{journal}{Physical Review Letters} \textbf{\bibinfo{volume}{56}},
  \bibinfo{pages}{900} (\bibinfo{year}{1986}),
  \urlprefix\url{http://link.aps.org/abstract/PRL/v56/p900}.

\bibitem[{\citenamefont{Lin}(1997)}]{Lin97}
\bibinfo{author}{\bibfnamefont{Q.~G.} \bibnamefont{Lin}},
  \bibinfo{journal}{Phys. Rev. A} \textbf{\bibinfo{volume}{56}},
  \bibinfo{pages}{1938} (\bibinfo{year}{1997}),
  \urlprefix\url{http://link.aps.org/abstract/PRA/v56/p1938}.

\bibitem[{\citenamefont{Dong et~al.}(1998)\citenamefont{Dong, Hou, and
  Ma}}]{Dong98}
\bibinfo{author}{\bibfnamefont{S.~H.} \bibnamefont{Dong}},
  \bibinfo{author}{\bibfnamefont{X.~W.} \bibnamefont{Hou}}, \bibnamefont{and}
  \bibinfo{author}{\bibfnamefont{Z.~Q.} \bibnamefont{Ma}},
  \bibinfo{journal}{Phys. Rev. A} \textbf{\bibinfo{volume}{58}},
  \bibinfo{pages}{2790} (\bibinfo{year}{1998}).

\bibitem[{\citenamefont{Sheka et~al.}(2003)\citenamefont{Sheka, Ivanov, and
  Mertens}}]{Sheka03}
\bibinfo{author}{\bibfnamefont{D.}~\bibnamefont{Sheka}},
  \bibinfo{author}{\bibfnamefont{B.}~\bibnamefont{Ivanov}}, \bibnamefont{and}
  \bibinfo{author}{\bibfnamefont{F.~G.} \bibnamefont{Mertens}},
  \bibinfo{journal}{Phys. Rev. A} \textbf{\bibinfo{volume}{68}},
  \bibinfo{eid}{012707} (pages~\bibinfo{numpages}{5}) (\bibinfo{year}{2003}),
  \urlprefix\url{http://link.aps.org/abstract/PRA/v68/e012707}.

\bibitem[{\citenamefont{Kobe and Liang}(1988)}]{Kobe88}
\bibinfo{author}{\bibfnamefont{D.~H.} \bibnamefont{Kobe}} \bibnamefont{and}
  \bibinfo{author}{\bibfnamefont{J.~Q.} \bibnamefont{Liang}},
  \bibinfo{journal}{Phys. Rev. A} \textbf{\bibinfo{volume}{37}},
  \bibinfo{pages}{1133} (\bibinfo{year}{1988}).

\bibitem[{\citenamefont{Decanini and Folacci}(2003)}]{Decanini03}
\bibinfo{author}{\bibfnamefont{Y.}~\bibnamefont{Decanini}} \bibnamefont{and}
  \bibinfo{author}{\bibfnamefont{A.}~\bibnamefont{Folacci}},
  \bibinfo{journal}{Phys. Rev. A} \textbf{\bibinfo{volume}{67}},
  \bibinfo{pages}{042704} (\bibinfo{year}{2003}).

\bibitem[{\citenamefont{Aharonov et~al.}(1984)\citenamefont{Aharonov, Au,
  Lerner, and Liang}}]{Aharonov84}
\bibinfo{author}{\bibfnamefont{Y.}~\bibnamefont{Aharonov}},
  \bibinfo{author}{\bibfnamefont{C.~K.} \bibnamefont{Au}},
  \bibinfo{author}{\bibfnamefont{E.~C.} \bibnamefont{Lerner}},
  \bibnamefont{and} \bibinfo{author}{\bibfnamefont{J.~Q.} \bibnamefont{Liang}},
  \bibinfo{journal}{Phys. Rev. D} \textbf{\bibinfo{volume}{29}},
  \bibinfo{pages}{2396} (\bibinfo{year}{1984}),
  \urlprefix\url{http://link.aps.org/abstract/PRD/v29/p2396}.

\bibitem[{\citenamefont{Henneberger}(1980)}]{Henneberger80}
\bibinfo{author}{\bibfnamefont{W.~C.} \bibnamefont{Henneberger}},
  \bibinfo{journal}{Phys. Rev. A} \textbf{\bibinfo{volume}{22}},
  \bibinfo{pages}{1383} (\bibinfo{year}{1980}).

\bibitem[{\citenamefont{Ruijsenaars}(1983)}]{Ruijsenaars83}
\bibinfo{author}{\bibfnamefont{S.~N.~M.} \bibnamefont{Ruijsenaars}},
  \bibinfo{journal}{Annals of Physics} \textbf{\bibinfo{volume}{146}},
  \bibinfo{pages}{1} (\bibinfo{year}{1983}),
  \urlprefix\url{http://www.sciencedirect.com/science/article/B6WB1-4DDR3KP-NJ%
/2/d8f56ee8b6ab36e8635946f3f839dfbf}.

\bibitem[{\citenamefont{Belavin and Polyakov}(1975)}]{Belavin75}
\bibinfo{author}{\bibfnamefont{A.~A.} \bibnamefont{Belavin}} \bibnamefont{and}
  \bibinfo{author}{\bibfnamefont{A.~M.} \bibnamefont{Polyakov}},
  \bibinfo{journal}{JETP Lett.} \textbf{\bibinfo{volume}{22}},
  \bibinfo{pages}{245} (\bibinfo{year}{1975}).

\bibitem[{\citenamefont{Ivanov et~al.}(1999{\natexlab{a}})\citenamefont{Ivanov,
  Murav'ev, and Sheka}}]{Ivanov99}
\bibinfo{author}{\bibfnamefont{B.~A.} \bibnamefont{Ivanov}},
  \bibinfo{author}{\bibfnamefont{V.~M.} \bibnamefont{Murav'ev}},
  \bibnamefont{and} \bibinfo{author}{\bibfnamefont{D.~D.} \bibnamefont{Sheka}},
  \bibinfo{journal}{Journal of Experimental and Theoretical Physics}
  \textbf{\bibinfo{volume}{89}}, \bibinfo{pages}{583}
  (\bibinfo{year}{1999}{\natexlab{a}}),
  \urlprefix\url{http://link.aip.org/link/?JET/89/583/1}.

\bibitem[{\citenamefont{Ivanov and Sheka}(2005)}]{Ivanov05b}
\bibinfo{author}{\bibfnamefont{B.~A.} \bibnamefont{Ivanov}} \bibnamefont{and}
  \bibinfo{author}{\bibfnamefont{D.~D.} \bibnamefont{Sheka}},
  \bibinfo{journal}{JETP Letters} \textbf{\bibinfo{volume}{82}},
  \bibinfo{pages}{436} (\bibinfo{year}{2005}),
  \urlprefix\url{http://link.aip.org/link/?JTP/82/436/1}.

\bibitem[{\citenamefont{Ivanov}(1995)}]{Ivanov95g}
\bibinfo{author}{\bibfnamefont{B.~A.} \bibnamefont{Ivanov}},
  \bibinfo{journal}{JETP Lett.} \textbf{\bibinfo{volume}{61}},
  \bibinfo{pages}{917} (\bibinfo{year}{1995}).

\bibitem[{\citenamefont{Rodriguez}(1989)}]{Rodriguez89}
\bibinfo{author}{\bibfnamefont{J.~P.} \bibnamefont{Rodriguez}},
  \bibinfo{journal}{Phys. Rev. B} \textbf{\bibinfo{volume}{39}},
  \bibinfo{pages}{2906} (\bibinfo{year}{1989}),
  \urlprefix\url{http://link.aps.org/abstract/PRB/v39/p2906}.

\bibitem[{\citenamefont{Pereira et~al.}(1995)\citenamefont{Pereira, Pires, and
  Gouv{\^e}a}}]{Pereira95}
\bibinfo{author}{\bibfnamefont{A.~R.} \bibnamefont{Pereira}},
  \bibinfo{author}{\bibfnamefont{A.~S.~T.} \bibnamefont{Pires}},
  \bibnamefont{and} \bibinfo{author}{\bibfnamefont{M.~E.}
  \bibnamefont{Gouv{\^e}a}}, \bibinfo{journal}{Phys. Rev. B}
  \textbf{\bibinfo{volume}{51}}, \bibinfo{pages}{15974} (\bibinfo{year}{1995}),
  \urlprefix\url{http://link.aps.org/abstract/PRB/v51/p15974}.

\bibitem[{\citenamefont{Ivanov et~al.}(1999{\natexlab{b}})\citenamefont{Ivanov,
  Muravyov, and Sheka}}]{Ivanov99a}
\bibinfo{author}{\bibfnamefont{B.~A.} \bibnamefont{Ivanov}},
  \bibinfo{author}{\bibfnamefont{V.~M.} \bibnamefont{Muravyov}},
  \bibnamefont{and} \bibinfo{author}{\bibfnamefont{D.~D.} \bibnamefont{Sheka}},
  \bibinfo{journal}{Ukr. Fiz. Zh.} \textbf{\bibinfo{volume}{44}},
  \bibinfo{pages}{1404} (\bibinfo{year}{1999}{\natexlab{b}}).

\bibitem[{\citenamefont{Feinberg}(1963)}]{Feinberg63}
\bibinfo{author}{\bibfnamefont{E.}~\bibnamefont{Feinberg}},
  \bibinfo{journal}{Sov. Phys. Usp.} \textbf{\bibinfo{volume}{5}},
  \bibinfo{pages}{753} (\bibinfo{year}{1963}),
  \urlprefix\url{http://www.ufn.ru/archive/russian/abstracts/abst3970.html}.

\end{thebibliography}

\end{document}